\documentclass{elsart}

\usepackage{amssymb}

\begin{document}

\begin{frontmatter}



\title{Heavy hadron spectroscopy and the bag model }


\author{A.~Bernotas\corauthref{cor}} and
\ead{bernotas@itpa.lt}
\author{V.~\v{S}imonis}
\ead{simonis@itpa.lt}
\corauth[cor]{Corresponding author.}

\address{Vilnius University Institute of Theoretical Physics and Astronomy, \newline 
A.~Go\v{s}tauto 12, LT-01108 Vilnius, Lithuania}

\begin{abstract}
Some time ago a slightly improved variant of bag model (the modified bag
model) suitable for the unified description of light and heavy hadrons was
developed. The main goal of the present work was to calculate the masses of
the ground state baryons containing bottom quarks in the framework of this
model. For completeness the predictions for other heavy hadrons are also
given. The reasonable agreement of our results with other theoretical
calculations and available experimental data suggests that our predictions
could serve as a useful complementary tool for the interpretation of heavy
hadron spectra.
\end{abstract}

\end{frontmatter}

\section{Introduction}

During the last decade a significant progress has been achieved in the
experimental and theoretical studies of heavy hadrons (for review see \cite
{01R07}). In the nearest future a considerable amount of new experimental
data in the bottom (beauty) sector is expected from the Large Hadron Collider
(LHC). Therefore the theoretical calculation of the masses of heavy baryons
containing bottom quarks becomes more urgent. Ideally, one would like to
obtain the hadron masses from the first principles, say, performing the
nonperturbative lattice QCD calculations. But the available quenched lattice
predictions for heavy hadron masses \cite
{02BeA96,03AeA00,04MLW02,05FMT03,06CH05} still suffer from rather large
numerical uncertainties due to finite size effects, systematic and
statistical errors. The further progress in lattice QCD is connected with
improved calculations which take account of degrees of freedom associated
with the light (up, down, and strange) sea quarks. Some preliminary results
there already exist \cite{07NG07}, and impressive success of the lattice QCD
in the heavy hadron sector is the prediction of mass of the $B_{c}$ meson 
\cite{08AeA05}. Despite the lack of high accuracy,
even the so far available lattice calculations are of crucial importance for the
consistent treatment of the heavy hadron properties. First of all, they
serve as a good starting point for the further analysis of
heavy hadron spectrum. On the other hand, if one is interested only in
calculating the mass splittings, then major uncertainties cancel out and the
accuracy of lattice predictions increases.

Of course, there are other methods available: QCD sum rules, heavy quark
effective theory, various potential models, bag model, etc. As a rule, they
are based on some assumptions and approximations. However, taken all together,
they have become a powerful tool capable to give reasonable predictions for the
heavy hadron properties. QCD sum rules is another (besides lattice QCD)
nonperturbative approach which could be applied to study heavy baryon
spectrum. Treatment based on this approach \cite{09DN07,10LCLHZ08} provides the
results consistent with experimental observations, yet one must not
expect very high precision in this case. Another very fruitful and
interesting approach is based on the expansion in inverse powers of the
heavy quark mass. The effective heavy quark theory obtained in this way has
become rather popular tool in heavy hadron physics and a new approximate heavy
quark symmetry has been discovered (for review see \cite
{11KKP94,12N94,13M97,14BSU97}). Various potential models are widely used and
generally regarded as being rather successful. Indeed, it is amazing
how well the potential model predictions fit large and varied data set. Even the
early treatments \cite{15CIK79,16GI85,17CI86} predict the masses of many
hadrons with good accuracy. Among the recent treatments there are extensive
developments of the traditional approach \cite{18PR08,19RP07}, calculations
using the method of hyperspherical harmonics \cite{20CDR98}, solving the
Faddeev equations \cite{21GVV07}, variational calculations \cite
{22AAHN04,23AAHN05,24AHNV07}, nonrelativistic calculations using
quark-diquark approximation \cite{25GKLO00,26KLPS02}, etc. Relativistic
effects are taken into account by building up the relativistic quark models
based on approximate solution of the Bethe--Salpeter equation \cite{27MMMP06}%
, or using quark-diquark approximation to simplify relativistic dynamics 
\cite{28EFGMS97,29EFGM02,30EFG05}. Almost all approaches give predictions in
fair agreement with data, at least in the charm sector. This would seem to
imply that there must be a truthful basis in all this machinery. However,
one should not forget that the potential models are essentially of a
phenomenological nature.
 The interaction built into a particular model is
in fact the effective interaction, therefore there must be no surprise that
sometimes different approaches give similar results and vice versa. One
possible and widely used form of the interaction, proposed in the seminal work 
\cite{31DGG75}, imitates the one-gluon exchange. Other examples of the
effective interactions used in the heavy hadron spectroscopy are: the
phenomenological extension of the instanton-induced force \cite{27MMMP06}
and the flavor dependent interactions suggested by the
Goldstone-boson exchange \cite{20CDR98,32GR96,33GPVW98}. Some mixture of
one-gluon-exchange and Goldstone-boson-exchange induced interactions is also
possible \cite{22AAHN04,34VFV05}. The bag model was originally designed for
the ultrarelativistic case of the light quarks \cite{35DJJK75}. The first
straightforward application of the model to calculate the spectrum of the
hadrons containing heavy quarks was not very successful \cite{36JK76}. Later
on, there were several attempts to calculate the spectrum of the heavy
hadrons using improved versions of the model \cite{37P79,38IDS82}. Another
improved version of the bag model suitable for the unified description of
light and heavy hadrons was proposed by present authors in the Ref.~\cite
{39BS04}. Now we present our predictions for the masses of the ground state
baryons containing bottom quarks that are calculated using this version of the model
(the modified bag model).

The paper is organized as follows: Section 2 very briefly describes the
modified bag model. For more details we refer to \cite{39BS04}. In the
beginning of Section 3 we present our previous calculations (with some minor
improvements)
 for the charm sector baryons and for the heavy mesons belonging to
the charm and bottom sectors. Our results are compared with the estimates
obtained in various other approaches and with experimental data where
available. This may be regarded as a test of the model and gives us some
feeling of what should be expected in the case of the bottom baryons. Then the
modified bag model predictions for the ground state baryons containing
bottom quarks are presented and discussed. This is the main result of the
present paper. Finally, in Section 4 we conclude with the considerations
on the role of the bag model in the hadron spectroscopy.

\section{Modified bag model}

The calculation of the ground state hadron mass proceeds in two steps.
First, the so-called bag energy associated with particular hadron is
calculated. It depends on the bag radius $R$ and is defined by 
\begin{equation}
E=E_{V}+E_{Q}+E_{M}\,+E_{C}\ ,  \label{eq2.01}
\end{equation}
where $E_{V}$ is the bag volume energy, 
\begin{equation}
E_{V}=\frac{4\pi }{3}BR^{3}\ ,  \label{eq2.02}
\end{equation}
and $E_{Q}$ is the sum of single-particle quark energies, 
\begin{equation}
E_{Q}=\sum\limits_{i}n_{i}\varepsilon _{i}\,.  \label{eq2.03}
\end{equation}
In order to determine the numerical values of eigenenergies $\varepsilon _{i}
$ the free Dirac equation is solved for each quark subject to linear
boundary condition that ensures the vanishing of all vector currents at the
bag surface. $E_{M}$ and $E_{C}$ are the color-magnetic and color-electric
(Coulomb) interaction energies. They can be calculated explicitly \cite
{39BS04}. The numerical value of the bag energy is obtained minimizing~(\ref
{eq2.01}) as a function of $R$.

The quarks in the bag are not in an eigenstate of the total momentum. A part
of the bag energy is spurious and comes from the motion of the
center-of-mass (c.m.m. problem). In order to obtain the mass of the hadron
it is necessary to incorporate the c.m.m. corrections, in some way. We
follow the Ref. \cite{40HM90} and assume the relation between the calculated
bag-model energy $E$ and the mass $M$ of particular hadron to be 
\begin{equation}
E=\int \mathrm{d}^{3}s\,\Phi _{P}^{2}(s)\,\sqrt{M^{2}+s^{2}}\,,  \label{eq2.04}
\end{equation}
where $\Phi _{P}(s)$ is a Gauss profile,
\begin{equation}
\Phi _{P}(s)=\left( \frac{3}{2\pi P^{2}}\right) ^{3/4}\,\exp \left( -\frac{%
3s^{2}}{4P^{2}}\right) \,.  \label{eq2.05}
\end{equation}

The effective momentum $P$ which specifies the momentum distribution is
defined by

\begin{equation}
P^{2}=\gamma \sum\limits_{i}n_{i}p_{i}^{2}\,,  \label{eq2.06}
\end{equation}
where $p_{i}$ are the momenta of the individual quarks and $\gamma $ is the
model parameter governing the c.m.m. correction.

In order to obtain the mass of the particle, Eq.~(\ref{eq2.04}) is to be
solved numerically. However, in the presence of bottom quarks this equation
is practically equivalent to the relation \cite{40HM90} 
\begin{equation}
M^{2}=E^{2}-P^{2}\,,  \label{eq2.07}
\end{equation}
and in this case we prefer to use this simple relation instead of the rather
cumbersome procedure based on Eq.~(\ref{eq2.04}).

The interaction energies are computed to the first order in the scale-%
dependent strong coupling constant $\alpha _{\mathrm{c}}(R)$, and for the
strange, charmed, and bottom quarks we use the running mass $\overline{m}%
_{f}(R)$ \cite{39BS04}. The light (up and down in our case) quarks are
assumed to be massless.

For the running coupling constant we use the expression proposed by Donoghue
and Johnson \cite{41DJ80} 
\begin{equation}
\alpha _{\mathrm{c}}(R)=\frac{2\pi }{9\ln (A+R_{0}/R)}\,,  \label{eq2.08}
\end{equation}
where $R_{0}$ is the scale parameter which plays the role similar to QCD
constant $\Lambda $. The model parameter $A$ serves to avoid divergences in
the case when $R\rightarrow R_{0}$.

For the mass function $\overline{m}_{f}(R)$ we use the expression
\begin{equation}
\overline{m}_{f}(R)=\widetilde{m}_{f}+\alpha _{\mathrm{c}}(R)\cdot \delta
_{f}\,,  \label{eq2.09}
\end{equation}
thus, for each quark flavor there are two free model parameters $%
\widetilde{m}_{f}$ and $\delta _{f}$.

Altogether we have ten free parameters in the model. The four of them ($B$, $%
\gamma $, $A$, and $R_{0}$) were determined by fitting calculated masses of
light hadrons ($N$, $\Delta $, $\pi $, and the average mass of the $\omega$%
--$\rho $ system) to experimental data. The numerical values of these
parameters are the same as in the Ref.~\cite{39BS04}: $B=7.597\cdot 10^{-4}~$%
GeV$^{4}$, $R_{0}=2.543~$GeV$^{-1}$, $A=1.070$, $\gamma =1.958$. In order to
fix the remaining six parameters ($\widetilde{m}_{s}$, $\delta _{s}$, $%
\widetilde{m}_{c}$, $\delta _{c}$, $\widetilde{m}_{b}$, $\delta _{b}$)
necessary to define the mass functions $\overline{m}_{f}(R)$ we have
employed the masses of vector mesons ($\phi ,$ $J/\psi ,$ $%
\Upsilon%
$) accompanied by the mass values of the lightest baryons $\Lambda _{f}$
belonging to the corresponding flavor sector. The results are presented in
Table~\ref{t2.1}.

\begin{table}[tbp] \centering%
\caption{Parameters (in GeV) used to determine the behavior of the mass 
functions  $\overline{m}_{f}(R)$.\label{t2.1}} 
\begin{tabular}{cccc}
\hline
$f$ & $s$ & $c$ & $b$ \\ \hline
$\widetilde{m}_{f}$ & 0.161 & 1.458 & 4.793 \\ 
$\delta _{f}$ & 0.156 & 0.112 & 0.061 \\ \hline
\end{tabular}
\end{table}%

The numerical values of the parameters $\widetilde{m}_{c}$, $\delta _{c}$, $%
\widetilde{m}_{b}$, and $\delta _{b}$ differ slightly from the corresponding
values adopted in \cite{39BS04} because in the present work we have used
new, more accurate mass values of $\Lambda _{c}$ (2.286$~$GeV) and $\Lambda
_{b}$ (5.620$~$GeV) \cite{42AeA08}.

We end up this section with the remark about the baryon mixing problem. It
is well known that the hyperfine interaction mixes the wave
functions of ground state spin-$1/2$ baryons containing three quarks of
different flavors \cite{44FLNC81}. In our case such baryons are $\Xi _{c}$
and $\Xi _{c}^{\prime }$, $\Xi _{b}$ and $\Xi _{b}^{\prime }$, $\Xi _{bc}$
and $\Xi _{bc}^{\prime }$, $\Omega _{bc}$ and $\Omega _{bc}^{\prime }$. In
order to avoid ambiguities associated with the ordering of the quarks in the
wave function $|(q_{1}q_{2})^{S}q_{3}\rangle $ we simply calculate the
off-diagonal elements of the interaction energy matrix with consequent matrix
diagonalization (for details see \cite{45BS08}). The alternative choice
would be to use the so-called optimal basis by picking up the heaviest quark
as the third one in the spin coupling scheme $(q_{1}q_{2})^{S}q_{3}$ \cite{44FLNC81,45BS08}. For
the c.m.m. uncorrected bag energies both choices give practically identical
results.
 The resulting mass values may differ slightly. This is so, because the
spurious c.m.m. energy, which must be subtracted from the hadron energy,
depends on the hadron bag radius $R_{H}$. When we are dealing with a mixed
set of wave functions, we minimize the trace of the energy matrix, $%
E_{B}+E_{B^{\prime }}$, which remains invariant under state mixing. So, in
this case the bag radii for both mixed hadrons coincide. On the other hand,
if the optimal basis is used, the natural choice would be to minimize the
energy of each baryon individually. Then the radii of baryons under
consideration are no longer identical (practically the difference is quite
small). The shift in mass caused by such change of bag radii does not exceed
2~MeV, which is obviously smaller than the systematic uncertainties of the
model and, in principle, has nothing to do with the baryon mixing.

\section{Predictions for the ground state hadron masses}

Strictly speaking, the bag model is not derivable from QCD and the quality
of its predictions is not quite clear. We can only compare the
results of calculations with experimental data (if available) as well as
with calculations using various other approaches. A good starting point for such comparison ought to be the charm sector. As a first step let us compare our predictions
for the hadrons from this sector with the results obtained in other variants
of the bag model. The results for hadron masses and mass splittings are
presented in Tables~\ref{t3.1} and \ref{t3.2} respectively. The
columns of the tables denoted as MIT contain the original results from the
Ref.~\cite{36JK76}. In Ref.~\cite{46SV98} the quark--quark interaction is
treated perturbatively. Moreover, in this work the nonphysical self-energy
term (which was present in the original MIT version of the model) is
omitted. In Ref.~\cite{37P79} the improvement of the hadronic mass spectrum
is achieved introducing some extra dependence of the bag energy on the
heaviest (inside the hadron) quark mass. The variant of model used in Ref.~%
\cite{38IDS82} was constructed for hadrons containing one heavy quark. In
this approach the heavy quark is treated as a point source of color fields
located at the center of the bag. The experimental values are taken
from the Particle Data Tables \cite{42AeA08}. For the isospin multiplets the
averaged values are used here and further on. To show the quark content of
the hadrons we use the symbols $b$, $c$, $s$ for the bottom, charmed, and
strange quarks, respectively. For the sake of simplicity the symbol $u$ is
used for both light (up or down) quarks. The corresponding antiquarks are
denoted as $\overline{b}$, $\overline{c}$, $\overline{s}$, and $\overline{u}$%
. In the last row of the Table~\ref{t3.1} the values of the root mean
squared deviation from the experimenal mass spectra

\[
\chi =\left[ \frac{1}{N}\sum\limits_{i=1}^{N}\left( M^{i}-M_{\mathrm{ex}%
}^{i}\right) ^{2}\right] ^{1/2} 
\]
are presented. The mass of the $\Xi _{cc}$ baryon has not been included in
the summation.

\begin{table}[tbp] \centering%
\caption{Masses  (in GeV) of ground state hadrons from the charm sector, calculated in 
five variants of the bag model as described in the text. The column denoted as Exp 
contains averaged over isomultiplets experimental data.\label{t3.1}} 
\begin{tabular}{llllllll}
\hline
Hadrons & Quarks & Exp & Our & MIT & \cite{46SV98} & \cite{37P79} & 
\cite{38IDS82} \\ \hline
\multicolumn{1}{c}{$J/\psi $} & \multicolumn{1}{c}{$c\overline{c}$} & 
\multicolumn{1}{c}{3.097} & \multicolumn{1}{c}{3.097} & \multicolumn{1}{c}{
3.095} & \multicolumn{1}{c}{3.15} & \multicolumn{1}{c}{3.095} & 
\multicolumn{1}{c}{---} \\ 
\multicolumn{1}{c}{$\eta _{c}$} & \multicolumn{1}{c}{$c\overline{c}$} & 
\multicolumn{1}{c}{2.980} & \multicolumn{1}{c}{3.005} & \multicolumn{1}{c}{
2.931} & \multicolumn{1}{c}{3.05} & \multicolumn{1}{c}{2.971} & 
\multicolumn{1}{c}{---} \\ 
\multicolumn{1}{c}{$D$} & \multicolumn{1}{c}{$c\overline{u}$} & 
\multicolumn{1}{c}{1.867} & \multicolumn{1}{c}{1.834} & \multicolumn{1}{c}{
1.726} & \multicolumn{1}{c}{1.82} & \multicolumn{1}{c}{1.800} & 
\multicolumn{1}{c}{1.83} \\ 
\multicolumn{1}{c}{$D^{*}$} & \multicolumn{1}{c}{$c\overline{u}$} & 
\multicolumn{1}{c}{2.008} & \multicolumn{1}{c}{2.002} & \multicolumn{1}{c}{
1.969} & \multicolumn{1}{c}{2.01} & \multicolumn{1}{c}{2.009} & 
\multicolumn{1}{c}{2.01} \\ 
\multicolumn{1}{c}{$D_{s}$} & \multicolumn{1}{c}{$c\overline{s}$} & 
\multicolumn{1}{c}{1.968} & \multicolumn{1}{c}{1.965} & \multicolumn{1}{c}{
1.885} & \multicolumn{1}{c}{1.98} & \multicolumn{1}{c}{1.957} & 
\multicolumn{1}{c}{1.92} \\ 
\multicolumn{1}{c}{$D_{s}^{*}$} & \multicolumn{1}{c}{$c\overline{s}$} & 
\multicolumn{1}{c}{2.112} & \multicolumn{1}{c}{2.119} & \multicolumn{1}{c}{
2.099} & \multicolumn{1}{c}{2.14} & \multicolumn{1}{c}{2.141} & 
\multicolumn{1}{c}{2.09} \\ 
\multicolumn{1}{c}{$\Lambda _{c}$} & \multicolumn{1}{c}{$cuu$} & 
\multicolumn{1}{c}{2.286} & \multicolumn{1}{c}{2.286} & \multicolumn{1}{c}{
2.214} & \multicolumn{1}{c}{2.29} & \multicolumn{1}{c}{2.243} & 
\multicolumn{1}{c}{2.28} \\ 
\multicolumn{1}{c}{$\Sigma _{c}$} & \multicolumn{1}{c}{$cuu$} & 
\multicolumn{1}{c}{2.454} & \multicolumn{1}{c}{2.393} & \multicolumn{1}{c}{
2.357} & \multicolumn{1}{c}{2.42} & \multicolumn{1}{c}{2.380} & 
\multicolumn{1}{c}{2.38} \\ 
\multicolumn{1}{c}{$\Sigma _{c}^{*}$} & \multicolumn{1}{c}{$cuu$} & 
\multicolumn{1}{c}{2.518} & \multicolumn{1}{c}{2.489} & \multicolumn{1}{c}{
2.461} & \multicolumn{1}{c}{2.53} & \multicolumn{1}{c}{2.481} & 
\multicolumn{1}{c}{2.49} \\ 
\multicolumn{1}{c}{$\Xi _{c}$} & \multicolumn{1}{c}{$csu$} & 
\multicolumn{1}{c}{2.469} & \multicolumn{1}{c}{2.468} & \multicolumn{1}{c}{
2.396} & \multicolumn{1}{c}{2.48} & \multicolumn{1}{c}{2.425} & 
\multicolumn{1}{c}{2.43} \\ 
\multicolumn{1}{c}{$\Xi _{c}^{\prime }$} & \multicolumn{1}{c}{$csu$} & 
\multicolumn{1}{c}{2.577} & \multicolumn{1}{c}{2.546} & \multicolumn{1}{c}{
2.507} & \multicolumn{1}{c}{---} & \multicolumn{1}{c}{2.530} & 
\multicolumn{1}{c}{2.50} \\ 
\multicolumn{1}{c}{$\Xi _{c}^{*}$} & \multicolumn{1}{c}{$csu$} & 
\multicolumn{1}{c}{2.646} & \multicolumn{1}{c}{2.638} & \multicolumn{1}{c}{
2.603} & \multicolumn{1}{c}{2.67} & \multicolumn{1}{c}{2.624} & 
\multicolumn{1}{c}{2.60} \\ 
\multicolumn{1}{c}{$\Omega _{c}$} & \multicolumn{1}{c}{$css$} & 
\multicolumn{1}{c}{2.697} & \multicolumn{1}{c}{2.697} & \multicolumn{1}{c}{
2.653} & \multicolumn{1}{c}{2.73} & \multicolumn{1}{c}{2.678} & 
\multicolumn{1}{c}{2.61} \\ 
\multicolumn{1}{c}{$\Omega _{c}^{*}$} & \multicolumn{1}{c}{$css$} & 
\multicolumn{1}{c}{2.768} & \multicolumn{1}{c}{2.783} & \multicolumn{1}{c}{
2.742} & \multicolumn{1}{c}{---} & \multicolumn{1}{c}{2.764} & 
\multicolumn{1}{c}{2.71} \\ 
\multicolumn{1}{c}{$\Xi _{cc}$} & \multicolumn{1}{c}{$ccu$} & 
\multicolumn{1}{c}{3.519} & \multicolumn{1}{c}{3.557} & \multicolumn{1}{c}{
3.538} & \multicolumn{1}{c}{3.66} & \multicolumn{1}{c}{3.511} & 
\multicolumn{1}{c}{---} \\ 
\multicolumn{1}{c}{$\Xi _{cc}^{*}$} & \multicolumn{1}{c}{$ccu$} & 
\multicolumn{1}{c}{---} & \multicolumn{1}{c}{3.661} & \multicolumn{1}{c}{
3.661} & \multicolumn{1}{c}{---} & \multicolumn{1}{c}{3.630} & 
\multicolumn{1}{c}{---} \\ 
\multicolumn{1}{c}{$\Omega _{cc}$} & \multicolumn{1}{c}{$ccs$} & 
\multicolumn{1}{c}{---} & \multicolumn{1}{c}{3.710} & \multicolumn{1}{c}{
3.690} & \multicolumn{1}{c}{3.82} & \multicolumn{1}{c}{3.664} & 
\multicolumn{1}{c}{---} \\ 
\multicolumn{1}{c}{$\Omega _{cc}^{*}$} & \multicolumn{1}{c}{$ccs$} & 
\multicolumn{1}{c}{---} & \multicolumn{1}{c}{3.800} & \multicolumn{1}{c}{
3.795} & \multicolumn{1}{c}{---} & \multicolumn{1}{c}{3.764} & 
\multicolumn{1}{c}{---} \\ 
\multicolumn{1}{c}{$\Omega _{ccc}$} & \multicolumn{1}{c}{$ccc$} & 
\multicolumn{1}{c}{---} & \multicolumn{1}{c}{4.777} & \multicolumn{1}{c}{
4.827} & \multicolumn{1}{c}{4.98} & \multicolumn{1}{c}{4.747} & 
\multicolumn{1}{c}{---} \\ \hline
\multicolumn{1}{c}{$\chi $} &  & \multicolumn{1}{c}{---} & 0.023 & 0.067 & 
\multicolumn{1}{c}{0.03} & \multicolumn{1}{c}{0.036} & \multicolumn{1}{c}{
0.05} \\ \hline
\end{tabular}
\end{table}%

\begin{table}[tbp] \centering%
\caption{Mass splittings (in GeV) of some hadrons from the charm sector, calculated in 
five variants of the bag model as described in the text. The column denoted as Exp 
contains experimental data.\label{t3.2}} 
\begin{tabular}{lllllll}
\hline
Hadrons & Exp & Our & MIT & \cite{46SV98} & \cite{37P79} & \cite{38IDS82} \\ \hline
$J/\psi -\eta _{c}$ & \multicolumn{1}{c}{0.117} & \multicolumn{1}{c}{0.092}
& \multicolumn{1}{c}{0.164} & \multicolumn{1}{c}{0.10} & \multicolumn{1}{c}{
0.124} & \multicolumn{1}{c}{---} \\ 
$D^{*}-D$ & \multicolumn{1}{c}{0.141} & \multicolumn{1}{c}{0.168} & 
\multicolumn{1}{c}{0.244} & \multicolumn{1}{c}{0.19} & \multicolumn{1}{c}{
0.209} & \multicolumn{1}{c}{0.18} \\ 
$D_{s}^{*}-D_{s}$ & \multicolumn{1}{c}{0.144} & \multicolumn{1}{c}{0.154} & 
\multicolumn{1}{c}{0.214} & \multicolumn{1}{c}{0.16} & \multicolumn{1}{c}{
0.184} & \multicolumn{1}{c}{0.17} \\ 
$\Sigma _{c}^{*}-\Sigma _{c}$ & \multicolumn{1}{c}{0.064} & 
\multicolumn{1}{c}{0.096} & \multicolumn{1}{c}{0.104} & \multicolumn{1}{c}{
0.11} & \multicolumn{1}{c}{0.101} & \multicolumn{1}{c}{0.11} \\ 
$\Xi _{c}^{*}-\Xi _{c}^{\prime }$ & \multicolumn{1}{c}{0.069} & 
\multicolumn{1}{c}{0.092} & \multicolumn{1}{c}{0.096} & \multicolumn{1}{c}{
---} & \multicolumn{1}{c}{0.094} & \multicolumn{1}{c}{0.10} \\ 
$\Omega _{c}^{*}-\Omega _{c}$ & \multicolumn{1}{c}{0.071} & 
\multicolumn{1}{c}{0.086} & \multicolumn{1}{c}{0.089} & \multicolumn{1}{c}{
---} & \multicolumn{1}{c}{0.086} & \multicolumn{1}{c}{0.10} \\ 
$\Xi _{cc}^{*}-\Xi _{cc}$ & \multicolumn{1}{c}{---} & \multicolumn{1}{c}{
0.104} & \multicolumn{1}{c}{0.123} & \multicolumn{1}{c}{---} & 
\multicolumn{1}{c}{0.119} & \multicolumn{1}{c}{---} \\ 
$\Omega _{cc}^{*}-\Omega _{cc}$ & \multicolumn{1}{c}{---} & 
\multicolumn{1}{c}{0.090} & \multicolumn{1}{c}{0.106} & \multicolumn{1}{c}{
---} & \multicolumn{1}{c}{0.100} & \multicolumn{1}{c}{---} \\ \hline
\end{tabular}
\end{table}%

We see from Table~\ref{t3.1} that the original MIT results are in
serious conflict with experimental data and discrepancies seem to be of systematic character. The improved variants \cite{37P79}, \cite{38IDS82} and \cite
{46SV98} give evidently more reasonable predictions. The agreement of our
predictions with available experimental data is rather good, although there 
are some discrepancies. The most serious problem common to almost all
variants of bag model seems to be the mass of $\Sigma _{c}$ baryon (its
analog $\Sigma $ in the sector of light hadrons was also problematic \cite
{39BS04}). One possible way to improve the description of these states would
be the inclusion of chiral (pionic) corrections \cite{49MBX81,50MT83,51S84},
however, such extension is beyond the scope of the present investigation.

Let us refer to the Table~\ref{t3.2}. There we compare our predictions for some
hadron mass splittings with experimental data and other bag model
calculations. For the baryons all variants of the bag model give similar
predictions about 30\% larger than needed. This may indicate that
the interaction strength 
for baryons in the bag model is somewhat
overestimated, even in the versions of the model where the running coupling
constant is used. In the meson sector our predictions for the $D^{*}-D$ and $%
D_{s}^{*}-D_{s}$ mass splittings are better than in other approaches, though
still somewhat larger than experimental values. For the $J/\psi -\eta _{c}$
we together with Ref.~\cite{46SV98} predict somewhat smaller than required
mass difference. In general, we see that the overall agreement of our
predictions with available experimental data is good ($\Sigma _{c}$ mass
being the exception). Moreover, as a rule, our predictions almost in all
cases are better than the predictions given by other variants of the bag
model.

It is also useful to compare the results of our bag model calculations with
the predictions obtained in other (including more elaborated) approaches.
Let us start with the meson sector. In Tables~\ref{t3.3} and~\ref{t3.4} we
compare our predictions for the masses of heavy mesons with the calculations
in four different variants of the potential model. We have chosen for this
the relativized quark model \cite{16GI85}, the model based on the
Bethe--Salpeter equation \cite{52ZVR95}, the relativistic treatment using
quasipotential approach \cite{53EFG06,54EFG03}, and one specific variant of
nonrelativistic potential model \cite{34VFV05} where the Goldstone-boson
exchanges are considered together with the one-gluon-exchange. The
experimental values are from the Particle Data Tables \cite{42AeA08} with
the exeption of the $\eta _{b}$ meson. The mass value of $\eta _{b}$ is
taken from \cite{43AeA08}.

\begin{table}[tbp] \centering%
\caption{Masses and mass splittings (in GeV) for mesons belonging to the charm sector, 
calculated in various approaches as described in the text.
 The column denoted as Exp contains experimental data.\label{t3.3}} 
\begin{tabular}{llcccccc}
\hline
Mesons & Quarks & Exp & Our & \cite{16GI85} & \cite{52ZVR95} & \cite{53EFG06,54EFG03} & 
\cite{34VFV05} \\ \hline
$J/\psi $ & $c\overline{c}$ & 3.097 & 3.097 & 3.10 & 3.10 & 3.096 & 3.097 \\ 
$\eta _{c}$ & $c\overline{c}$ & 2.980 & 3.005 & 2.97 & 3.00 & 2.979 & 2.990
\\ 
$D$ & $c\overline{u}$ & 1.867 & 1.834 & 1.88 & 1.85 & 1.872 & 1.883 \\ 
$D^{*}$ & $c\overline{u}$ & 2.008 & 2.002 & 2.04 & 2.02 & 2.009 & 2.010 \\ 
$D_{s}$ & $c\overline{s}$ & 1.968 & 1.965 & 1.98 & 1.94 & 1.967 & 1.981 \\ 
$D_{s}^{*}$ & $c\overline{s}$ & 2.112 & 2.119 & 2.13 & 2.13 & 2.112 & 2.112
\\
$\chi $ &  & --- & 0.017 & 0.02 & 0.02 & 0.002 & 0.009 \\ 
$J/\psi -\eta _{c}$ & $c\overline{c}$ & 0.117 & 0.092 & 0.13 & 0.10 & 0.117
& 0.107 \\ 
$D^{*}-D$ & $c\overline{u}$ & 0.141 & 0.168 & 0.16 & 0.17 & 0.137 & 0.127 \\ 
$D_{s}^{*}-D_{s}$ & $c\overline{s}$ & 0.144 & 0.154 & 0.15 & 0.19 & 0.145 & 
0.131 \\ \hline
\end{tabular}
\end{table}%

\begin{table}[tbp] \centering%
\caption{Masses and mass splittings (in GeV) for mesons belonging to the bottom sector, 
calculated in various approaches as described in the text.
 The column denoted as Exp contains experimental data.\label{t3.4}} 
\begin{tabular}{cccccccc}
\hline
Mesons & Quarks & Exp & Our & \cite{16GI85} & \cite{52ZVR95} & \cite{53EFG06,54EFG03} & 
\cite{34VFV05} \\ \hline
\multicolumn{1}{l}{$%
\Upsilon%
$} & \multicolumn{1}{l}{$b\overline{b}$} & 9.460 & 9.460 & 9.46 & 9.46 & 
9.460 & 9.505 \\ 
\multicolumn{1}{l}{$\eta _{b}$} & \multicolumn{1}{l}{$b\overline{b}$} & 9.389
& 9.438 & 9.40 & 9.41 & 9.400 & 9.454 \\ 
\multicolumn{1}{l}{$B$} & \multicolumn{1}{l}{$b\overline{u}$} & 5.279 & 5.249
& 5.31 & 5.28 & 5.275 & 5.281 \\ 
\multicolumn{1}{l}{$B^{*}$} & \multicolumn{1}{l}{$b\overline{u}$} & 5.325 & 
5.306 & 5.37 & 5.33 & 5.326 & 5.321 \\ 
\multicolumn{1}{l}{$B_{s}$} & \multicolumn{1}{l}{$b\overline{s}$} & 5.367 & 
5.383 & 5.39 & 5.37 & 5.362 & 5.355 \\ 
\multicolumn{1}{l}{$B_{s}^{*}$} & \multicolumn{1}{l}{$b\overline{s}$} & 5.413
& 5.436 & 5.45 & 5.43 & 5.414 & 5.400 \\ 
\multicolumn{1}{l}{$B_{c}$} & \multicolumn{1}{l}{$b\overline{c}$} & 6.276 & 
6.304 & 6.27 & 6.26 & 6.270 & 6.277 \\ 
\multicolumn{1}{l}{$B_{c}^{*}$} & \multicolumn{1}{l}{$b\overline{c}$} & ---
& 6.342 & 6.34 & 6.34 & 6.332 & --- \\
\multicolumn{1}{l}{$\chi $} &  & --- & 0.027 & 0.03 & 0.01 & 0.005 & 0.031
\\ 
\multicolumn{1}{l}{$%
\Upsilon%
-\eta _{b}$} & \multicolumn{1}{l}{$b\overline{b}$} & 0.071 & 0.022 & 0.06 & 
0.05 & 0.060 & 0.051 \\ 
\multicolumn{1}{l}{$B^{*}-B$} & \multicolumn{1}{l}{$b\overline{u}$} & 0.046
& 0.057 & 0.06 & 0.05 & 0.051 & 0.040 \\ 
\multicolumn{1}{l}{$B_{s}^{*}-B_{s}$} & \multicolumn{1}{l}{$b\overline{s}$}
& 0.046 & 0.053 & 0.06 & 0.06 & 0.052 & 0.045 \\ 
\multicolumn{1}{l}{$B_{c}^{*}-B_{c}$} & \multicolumn{1}{l}{$b\overline{c}$}
& --- & 0.038 & 0.07 & 0.08 & 0.062 & --- \\ \hline
\end{tabular}
\end{table}%

As can be seen from these tables, the overall agreement of the heavy meson
spectrum calculated in our work with the experimental data is quite good.
Inspecting the meson mass differences we see that for the $D$ and $B$ mesons
consisting of one heavy (charmed or bottom) and one light (up or down) quark
the mass splitting obtained in our work is $\sim$20\% too large. For the $D_{s}$
and $B_{s}$ mesons consisting of one heavy and one strange quark the
agreement with experiment is better, while the predicted mass splitting is still
too large. On the other hand, for the $c\overline{c}$ system the mass
difference of $J/\psi$ and $\eta _{c}$ predicted in our model is about
30\% too small, and we expect similar discrepancy for the $%
\Upsilon%
-\eta _{b}$. Regrettably, the discrepancy with experiment in this case is
more severe. Our result is approximately three times smaller than
experimental value, and this seems to be the serious drawback of the model.
Since the $B_{c}$ meson is made of two heavy quarks, we expect
that our prediction for the $B_{c}^{*}-B_{c}$ mass difference should be
somewhat too small as well. So far, the $B_{c}$ meson is the only well-%
established system containing two different heavy quarks, and it naturally
has attracted much attention these years. In Table~\ref{t3.5} we
have collected a number of predictions made by various authors for the
ground state mass values of the $B_{c}$ and $B_{c}^{*}$ mesons.

\begin{table}[tbp] \centering%
\caption{Comparison of various predictions for the masses (in GeV) of
$B_{c}$ and $B_{c}^{*}$ mesons.\label{t3.5}} 
\begin{tabular}{ccccccccc}
\hline
& \cite{55GJ96} & \cite{56GKLT95} & \cite{57EQ94} & \cite{54EFG03} & \cite
{58G04} & \cite{59F99} & Our & \cite{08AeA05} \\ \hline
\multicolumn{1}{l}{$B_{c}$} & 6.247 & 6.253 & 6.264 & 6.270 & 6.271 & 6.286
& 6.304 & 6.304 \\ 
\multicolumn{1}{l}{$B_{c}^{*}$} & 6.308 & 6.317 & 6.337 & 6.332 & 6.338 & 
6.341 & 6.342 & --- \\ 
\multicolumn{1}{l}{$B_{c}^{*}-B_{c}$} & 0.061 & 0.064 & 0.073 & 0.062 & 0.067
& 0.055 & 0.038 & --- \\ \hline
\end{tabular}
\end{table}%

Except for the lattice QCD prediction \cite{08AeA05} given in the
last column of this table and our estimate before it, all others are the potential model
calculations. The numerical values of $B_{c}$ mass obtained in each model
depend on the particular potential and vary in the interval from 6.24 to
6.29~GeV. Our prediction lies slightly higher and so does the
lattice value. A possible reason for the mass of $B_{c}$ meson calculated in
our model to be slightly higher than potential model predictions may be a
supposed underestimate (look at the last row of Table~\ref{t3.5}) of the
interaction strength for the heavy-heavy hadrons. Recent experimental
results $6.276(\pm 6)$~GeV \cite{60AeA08} and $6.300(\pm 19)$~GeV \cite
{61AeA08} cover the range from 6.270 to 6.319~GeV and agree well with the
theoretical estimates. The discovery of the $B_{c}$ meson and almost precise
theoretical prediction of its mass is undoubtedly a great success of
experiment and theory.

Let us proceed to the spectra of heavy baryons. We compare our predictions
obtained in the modified bag model for baryons from the charm sector
(Table~\ref{t3.6}) and for baryons from the bottom sector (Tables~\ref{t3.7}
and~\ref{t3.8}) with some other estimates and experimental data. The works
we want to compare our results with are: the baryon mass estimates in
relativistic \cite{29EFGM02,30EFG05} and nonrelativistic \cite
{25GKLO00,26KLPS02} potential models with assumed quark-diquark ansatz,
usual nonrelativistic potential model \cite{19RP07}, variational
calculations \cite{23AAHN05,24AHNV07}, and estimates obtained using various
sum rules \cite{62LRP96}. All but the one experimental masses for bottom
baryons are from the Particle Data Tables \cite{42AeA08}. The mass value for 
$\Omega _{b}$ is taken from Ref.~\cite{65AeA08}.

\begin{table}[tbp] \centering%
\caption{Masses and mass splittings (in GeV) of the charm sector baryons, 
calculated in modified bag model and in other approaches as described in the text.
 The column denoted as Exp contains experimental data.\label{t3.6}}

\begin{tabular}{lcccccccc}
\hline
Baryons & Quarks & Exp & Our & \cite{29EFGM02,30EFG05} & \cite{25GKLO00,26KLPS02} & \cite
{19RP07} & \cite{23AAHN05,24AHNV07} & \cite{62LRP96} \\ \hline
$\Lambda _{c}$ & \multicolumn{1}{l}{$cuu$} & 2.286 & 2.286 & 2.297 & --- & 
2.268 & 2.295 & 2.285 \\ 
$\Sigma _{c}$ & \multicolumn{1}{l}{$cuu$} & 2.454 & 2.393 & 2.439 & --- & 
2.455 & 2.469 & 2.453 \\ 
$\Sigma _{c}^{*}$ & \multicolumn{1}{l}{$cuu$} & 2.518 & 2.489 & 2.518 & ---
& 2.519 & 2.548 & 2.530 \\ 
$\Xi _{c}$ & \multicolumn{1}{l}{$csu$} & 2.469 & 2.468 & 2.481 & --- & 
2.466 & 2.474 & 2.468 \\ 
$\Xi _{c}^{\prime }$ & \multicolumn{1}{l}{$csu$} & 2.577 & 2.546 & 2.578 & 
--- & 2.594 & 2.578 & 2.582 \\ 
$\Xi _{c}^{*}$ & \multicolumn{1}{l}{$csu$} & 2.646 & 2.638 & 2.654 & --- & 
2.649 & 2.655 & 2.651 \\ 
$\Omega _{c}$ & \multicolumn{1}{l}{$css$} & 2.697 & 2.697 & 2.698 & --- & 
2.718 & 2.681 & 2.710 \\ 
$\Omega _{c}^{*}$ & \multicolumn{1}{l}{$css$} & 2.768 & 2.783 & 2.768 & ---
& 2.776 & 2.755 & 2.775 \\ 
$\Xi _{cc}$ & \multicolumn{1}{l}{$ccu$} & 3.519 & 3.557 & 3.620 & 3.478 & 
3.676 & 3.612 & 3.676 \\ 
$\Xi _{cc}^{*}$ & \multicolumn{1}{l}{$ccu$} & --- & 3.661 & 3.727 & 3.610 & 
3.753 & 3.706 & 3.746 \\ 
$\Omega _{cc}$ & \multicolumn{1}{l}{$ccs$} & --- & 3.710 & 3.778 & 3.594 & 
3.815 & 3.702 & 3.787 \\ 
$\Omega _{cc}^{*}$ & \multicolumn{1}{l}{$ccs$} & --- & 3.800 & 3.872 & 3.730
& 3.876 & 3.783 & 3.851 \\ 
$\Omega _{ccc}$ & \multicolumn{1}{l}{$ccc$} & --- & 4.777 & --- & --- & 4.965
& --- & --- \\
$\chi $ &  & --- & 0.027 & 0.008 & --- & 0.012 & 0.015 & 0.007 \\  
$\Sigma _{c}^{*}-\Sigma _{c}$ & \multicolumn{1}{l}{$cuu$} & 0.064 & 0.096
& 0.079 & --- & 0.064 & 0.079 & 0.077 \\ 
$\Xi _{c}^{*}-\Xi _{c}^{\prime }$ & \multicolumn{1}{l}{$csu$} & 0.069 & 
0.092 & 0.076 & --- & 0.055 & 0.077 & 0.069 \\ 
$\Omega _{c}^{*}-\Omega _{c}$ & \multicolumn{1}{l}{$css$} & 0.071 & 0.086 & 
0.070 & --- & 0.058 & 0.074 & 0.065 \\ 
$\Xi _{cc}^{*}-\Xi _{cc}$ & \multicolumn{1}{l}{$ccu$} & --- & 0.104 & 0.107
& 0.132 & 0.077 & 0.094 & 0.070 \\ 
$\Omega _{cc}^{*}-\Omega _{cc}$ & \multicolumn{1}{l}{$ccs$} & --- & 0.090 & 
0.094 & 0.136 & 0.061 & 0.081 & 0.064 \\ \hline
\end{tabular}
\end{table}%

\begin{table}[tbp] \centering%
\caption{Masses (in GeV) of the bottom sector baryons, 
calculated in modified bag model and in other approaches as described in the text.
 The column denoted as Exp contains experimental data.\label{t3.7}} 
\begin{tabular}{ccccccccc}
\hline
Baryons & Quarks & Exp & Our & \cite{29EFGM02,30EFG05} & \cite{25GKLO00,26KLPS02} & \cite
{19RP07} & \cite{23AAHN05,24AHNV07} & \cite{62LRP96} \\ \hline
\multicolumn{1}{l}{$\Lambda _{b}$} & \multicolumn{1}{l}{$buu$} & 5.620 & 
\multicolumn{1}{r}{5.620} & \multicolumn{1}{r}{5.622} & \multicolumn{1}{r}{
---} & \multicolumn{1}{r}{5.612} & \multicolumn{1}{r}{5.643} & 
\multicolumn{1}{r}{5.627} \\ 
\multicolumn{1}{l}{$\Sigma _{b}$} & \multicolumn{1}{l}{$buu$} & 5.811 & 
\multicolumn{1}{r}{5.755} & \multicolumn{1}{r}{5.805} & \multicolumn{1}{r}{
---} & \multicolumn{1}{r}{5.833} & \multicolumn{1}{r}{5.851} & 
\multicolumn{1}{r}{5.818} \\ 
\multicolumn{1}{l}{$\Sigma _{b}^{*}$} & \multicolumn{1}{l}{$buu$} & 5.833
& \multicolumn{1}{r}{5.787} & \multicolumn{1}{r}{5.834} & \multicolumn{1}{r}{
---} & \multicolumn{1}{r}{5.858} & \multicolumn{1}{r}{5.882} & 
\multicolumn{1}{r}{5.843} \\ 
\multicolumn{1}{l}{$\Xi _{b}$} & \multicolumn{1}{l}{$bsu$} & 5.792 & 
\multicolumn{1}{r}{5.809} & \multicolumn{1}{r}{5.812} & \multicolumn{1}{r}{
---} & \multicolumn{1}{r}{5.806} & \multicolumn{1}{r}{5.808} & 
\multicolumn{1}{r}{---} \\ 
\multicolumn{1}{l}{$\Xi _{b}^{\prime }$} & \multicolumn{1}{l}{$bsu$} & 
\multicolumn{1}{r}{---} & \multicolumn{1}{r}{5.911} & \multicolumn{1}{r}{
5.937} & \multicolumn{1}{r}{---} & \multicolumn{1}{r}{5.970} & 
\multicolumn{1}{r}{5.946} & \multicolumn{1}{r}{5.955} \\ 
\multicolumn{1}{l}{$\Xi _{b}^{*}$} & \multicolumn{1}{l}{$bsu$} & 
\multicolumn{1}{r}{---} & \multicolumn{1}{r}{5.944} & \multicolumn{1}{r}{
5.963} & \multicolumn{1}{r}{---} & \multicolumn{1}{r}{5.980} & 
\multicolumn{1}{r}{5.975} & \multicolumn{1}{r}{5.984} \\ 
\multicolumn{1}{l}{$\Omega _{b}$} & \multicolumn{1}{l}{$bss$} & 
\multicolumn{1}{r}{6.165} & \multicolumn{1}{r}{6.067} & \multicolumn{1}{r}{
6.065} & \multicolumn{1}{r}{---} & \multicolumn{1}{r}{6.081} & 
\multicolumn{1}{r}{6.033} & \multicolumn{1}{r}{6.075} \\ 
\multicolumn{1}{l}{$\Omega _{b}^{*}$} & \multicolumn{1}{l}{$bss$} & 
\multicolumn{1}{r}{---} & \multicolumn{1}{r}{6.096} & \multicolumn{1}{r}{
6.088} & \multicolumn{1}{r}{---} & \multicolumn{1}{r}{6.102} & 
\multicolumn{1}{r}{6.063} & \multicolumn{1}{r}{6.098} \\ 
\multicolumn{1}{l}{$\Xi _{bc}$} & \multicolumn{1}{l}{$bcu$} & 
\multicolumn{1}{r}{---} & \multicolumn{1}{r}{6.846} & \multicolumn{1}{r}{
6.933} & \multicolumn{1}{r}{6.82} & \multicolumn{1}{r}{7.011} & 
\multicolumn{1}{r}{6.919} & \multicolumn{1}{r}{7.029} \\ 
\multicolumn{1}{l}{$\Xi _{bc}^{\prime }$} & \multicolumn{1}{l}{$bcu$} & 
\multicolumn{1}{r}{---} & \multicolumn{1}{r}{6.891} & \multicolumn{1}{r}{
6.963} & \multicolumn{1}{r}{6.85} & \multicolumn{1}{r}{7.047} & 
\multicolumn{1}{r}{6.948} & \multicolumn{1}{r}{7.053} \\ 
\multicolumn{1}{l}{$\Xi _{bc}^{*}$} & \multicolumn{1}{l}{$bcu$} & 
\multicolumn{1}{r}{---} & \multicolumn{1}{r}{6.919} & \multicolumn{1}{r}{
6.980} & \multicolumn{1}{r}{6.90} & \multicolumn{1}{r}{7.074} & 
\multicolumn{1}{r}{6.986} & \multicolumn{1}{r}{7.083} \\ 
\multicolumn{1}{l}{$\Omega _{bc}$} & \multicolumn{1}{l}{$bcs$} & 
\multicolumn{1}{r}{---} & \multicolumn{1}{r}{6.999} & \multicolumn{1}{r}{
7.088} & \multicolumn{1}{r}{6.93} & \multicolumn{1}{r}{7.136} & 
\multicolumn{1}{r}{6.986} & \multicolumn{1}{r}{7.126} \\ 
\multicolumn{1}{l}{$\Omega _{bc}^{\prime }$} & \multicolumn{1}{l}{$bcs$} & 
\multicolumn{1}{r}{---} & \multicolumn{1}{r}{7.036} & \multicolumn{1}{r}{
7.116} & \multicolumn{1}{r}{6.97} & \multicolumn{1}{r}{7.165} & 
\multicolumn{1}{r}{7.009} & \multicolumn{1}{r}{7.148} \\ 
\multicolumn{1}{l}{$\Omega _{bc}^{*}$} & \multicolumn{1}{l}{$bcs$} & 
\multicolumn{1}{r}{---} & \multicolumn{1}{r}{7.063} & \multicolumn{1}{r}{
7.130} & \multicolumn{1}{r}{7.00} & \multicolumn{1}{r}{7.187} & 
\multicolumn{1}{r}{7.046} & \multicolumn{1}{r}{7.165} \\ 
\multicolumn{1}{l}{$\Omega _{bcc}$} & \multicolumn{1}{l}{$bcc$} & 
\multicolumn{1}{r}{---} & \multicolumn{1}{r}{7.984} & \multicolumn{1}{r}{---}
& \multicolumn{1}{r}{---} & \multicolumn{1}{r}{8.245} & \multicolumn{1}{r}{
---} & \multicolumn{1}{r}{---} \\ 
\multicolumn{1}{l}{$\Omega _{bcc}^{*}$} & \multicolumn{1}{l}{$bcc$} & 
\multicolumn{1}{r}{---} & \multicolumn{1}{r}{8.005} & \multicolumn{1}{r}{---}
& \multicolumn{1}{r}{---} & \multicolumn{1}{r}{8.265} & \multicolumn{1}{r}{
---} & \multicolumn{1}{r}{---} \\ 
\multicolumn{1}{l}{$\Xi _{bb}$} & \multicolumn{1}{l}{$bbu$} & 
\multicolumn{1}{r}{---} & \multicolumn{1}{r}{10.062} & \multicolumn{1}{r}{
10.202} & \multicolumn{1}{r}{10.093} & \multicolumn{1}{r}{10.340} & 
\multicolumn{1}{r}{10.197} & \multicolumn{1}{r}{---} \\ 
\multicolumn{1}{l}{$\Xi _{bb}^{*}$} & \multicolumn{1}{l}{$bbu$} & 
\multicolumn{1}{r}{---} & \multicolumn{1}{r}{10.101} & \multicolumn{1}{r}{
10.237} & \multicolumn{1}{r}{10.133} & \multicolumn{1}{r}{10.367} & 
\multicolumn{1}{r}{10.236} & \multicolumn{1}{r}{10.398} \\ 
\multicolumn{1}{l}{$\Omega _{bb}$} & \multicolumn{1}{l}{$bbs$} & 
\multicolumn{1}{r}{---} & \multicolumn{1}{r}{10.208} & \multicolumn{1}{r}{
10.359} & \multicolumn{1}{r}{10.210} & \multicolumn{1}{r}{10.454} & 
\multicolumn{1}{r}{10.260} & \multicolumn{1}{r}{---} \\ 
\multicolumn{1}{l}{$\Omega _{bb}^{*}$} & \multicolumn{1}{l}{$bbs$} & 
\multicolumn{1}{r}{---} & \multicolumn{1}{r}{10.244} & \multicolumn{1}{r}{
10.389} & \multicolumn{1}{r}{10.257} & \multicolumn{1}{r}{10.486} & 
\multicolumn{1}{r}{10.297} & \multicolumn{1}{r}{10.483} \\ 
\multicolumn{1}{l}{$\Omega _{bbc}$} & \multicolumn{1}{l}{$bbc$} & 
\multicolumn{1}{r}{---} & \multicolumn{1}{r}{11.139} & \multicolumn{1}{r}{---
} & \multicolumn{1}{r}{11.12} & \multicolumn{1}{r}{11.535} & 
\multicolumn{1}{r}{---} & \multicolumn{1}{r}{---} \\ 
\multicolumn{1}{l}{$\Omega _{bbc}^{*}$} & \multicolumn{1}{l}{$bbc$} & 
\multicolumn{1}{r}{---} & \multicolumn{1}{r}{11.163} & \multicolumn{1}{r}{---
} & \multicolumn{1}{r}{11.18} & \multicolumn{1}{r}{11.554} & 
\multicolumn{1}{r}{---} & \multicolumn{1}{r}{---} \\ 
\multicolumn{1}{l}{$\Omega _{bbb}$} & \multicolumn{1}{l}{$bbb$} & 
\multicolumn{1}{r}{---} & \multicolumn{1}{r}{14.276} & \multicolumn{1}{r}{---
} & \multicolumn{1}{r}{---} & \multicolumn{1}{r}{14.834} & 
\multicolumn{1}{r}{---} & \multicolumn{1}{r}{---} \\ \hline
\end{tabular}
\end{table}%

\begin{table}[tbp] \centering%
\caption{Mass splittings (in GeV) of the bottom sector baryons, 
calculated in modified bag model and in other approaches
 as described in the text.\label{t3.8}} 
\begin{tabular}{lcccccc}
\hline
Baryons & Our & \cite{29EFGM02,30EFG05} & \cite{25GKLO00,26KLPS02} & \cite{19RP07} & 
\cite{23AAHN05,24AHNV07} & \cite{62LRP96} \\ \hline
$\Sigma _{b}^{*}-\Sigma _{b}$ & \multicolumn{1}{l}{0.032} & 0.029 & --- & 
0.025 & 0.031 & 0.025 \\ 
$\Xi _{b}^{*}-\Xi _{b}^{\prime }$ & \multicolumn{1}{l}{0.033} & 0.026 & ---
& 0.010 & 0.029 & 0.029 \\ 
$\Omega _{b}^{*}-\Omega _{b}$ & \multicolumn{1}{l}{0.029} & 0.023 & --- & 
0.021 & 0.030 & 0.023 \\ 
$\Xi _{bc}^{*}-\Xi _{bc}^{\prime }$ & \multicolumn{1}{l}{0.028} & 0.017 & 
0.050 & 0.027 & 0.038 & 0.030 \\ 
$\Omega _{bc}^{*}-\Omega _{bc}^{\prime }$ & \multicolumn{1}{l}{0.027} & 0.014
& 0.030 & 0.022 & 0.037 & 0.017 \\ 
$\Omega _{bcc}^{*}-\Omega _{bcc}$ & \multicolumn{1}{l}{0.021} & --- & --- & 
0.020 & --- & --- \\ 
$\Xi _{bb}^{*}-\Xi _{bb}$ & \multicolumn{1}{l}{0.039} & 0.035 & 0.040 & 0.027
& 0.039 & --- \\ 
$\Omega _{bb}^{*}-\Omega _{bb}$ & \multicolumn{1}{l}{0.036} & 0.030 & 0.047
& 0.032 & 0.037 & --- \\ 
$\Omega _{bbc}^{*}-\Omega _{bbc}$ & \multicolumn{1}{l}{0.024} & --- & 0.060
& 0.019 & --- & --- \\ \hline
\end{tabular}
\end{table}%

From Table~\ref{t3.6} we see that for baryons with one heavy quark ($\Xi
_{c}$ and $\Omega _{c}$ families) the predictions obtained in all approaches are
in good agreement with experimental observations. The bag model predicts
somewhat larger mass splittings for the states $\Xi _{c}^{*}-\Xi
_{c}^{\prime }$ and $\Omega _{c}^{*}-\Omega _{c}$, but this does not spoil
the fit substantially. For the corresponding baryons from the bottom sector
presented in Table~\ref{t3.7}, our predictions are in agreement with almost
all other calculations again. So far there are only five bottom baryons
observed ($\Lambda _{b}$, $\Sigma _{b}$, $\Sigma _{b}^{*}$, $\Xi _{b}$, and $%
\Omega _{b}$). We see that all predictions for $\Xi _{b}$ compare favourably with the
experimental data. For $\Sigma _{c}$, $\Sigma _{c}^{*}$ and $\Sigma _{b}$, $%
\Sigma _{b}^{*}$ baryons the bag model unfortunately is a bad adviser, while
the potential model calculations fit well the experimental data again. The experimental mass value of the doubly strange $b$~baryon $%
\Omega _{b}$ lies $\sim 100$~MeV higher than expected from the theoretical
estimates. All calculations in this case give similar results, and therefore
it is hard to understand such discrepancy. The only thing we can say is that
the $\Omega _{b}$ mass value observed in \cite{65AeA08} still needs
additional confirmation.

Another serious test of any model should be the ability to predict the
masses of the baryons containing two heavy quarks. The only available
candidate for this so far is the $\Xi _{cc}$ baryon, although the situation
is not quite clear. The SELEX Collaboration has reported the observation of $%
\Xi _{cc}$ in two different experiments \cite{66MeA02,67OeA05}. On the other
hand, \textsl{BABAR} \cite{68AeA06} and Belle \cite{69CeA06} collaborations found no evidence of such
baryon in their searches. From Table~\ref{t3.6} we see that only two
calculations (\cite{25GKLO00} and ours) predict $\Xi _{cc}$ mass compatible
with the SELEX result. All other approaches predict the mass of this baryon
to be 100--150~MeV higher. If we accept the SELEX result as true (regardless of 
the absence of independent confirmations), this would be the second
doubly-heavy hadron observed (the first was $B_{c}$ meson). In both cases
the predictions of the modified bag model are sufficiently good. The
comparison of our results with others shows that for doubly-heavy baryons
all approaches give qualitatively similar (the same ordering of states)
predictions, while the numerical values could differ substantially. The
lowest mass values are obtained in Refs.~\cite{25GKLO00,26KLPS02}, our
values lie higher by about 70--80~MeV, and all other approaches predict even
larger masses of these baryons. As regards the mass splittings of
doubly-heavy baryons $\Xi _{cc}^{*}-\Xi _{cc}$ and $\Omega _{cc}^{*}-\Omega
_{cc}$, the predictions vary from about 70 MeV \cite{62LRP96} to 130 MeV 
\cite{26KLPS02}, our estimate (approximately 100 MeV) together with the
predictions given by Ref.~\cite{29EFGM02} being somewhere in the middle. For
the triple-heavy baryon $\Omega _{ccc}$ the bag model predicted mass is
about 200 MeV smaller than the corresponding mass value in nonrelativistic
potential model ~\cite{19RP07}.

We expect the predicted spectrum of the bottom sector baryons to be
qualitatively similar to the spectrum in the charm sector, and indeed we
find similar regularities in the spectrum of bottom baryons. For the
doubly-heavy baryons containing heavy quarks of distinct flavor (charm and
bottom) the lowest baryon masses are predicted in Refs.~\cite
{25GKLO00,26KLPS02}. For the baryons from the $\Xi _{bc}$ family our
predictions are similar to \cite{25GKLO00,26KLPS02}, while the predictions of
Refs.~\cite{29EFGM02} and \cite{24AHNV07} are approximately 80~MeV higher
than ours, and predictions of Refs.~\cite{19RP07} and \cite{62LRP96} exceed
ours by about 170~MeV. For the $\Omega _{bc}$ family our results are similar
to the calculations of Ref.~\cite{24AHNV07} and about 60 MeV above the
values obtained in \cite{25GKLO00,26KLPS02}. The largest masses
(approximately 120~MeV higher than ours) are again given by \cite{19RP07}
and \cite{62LRP96}. For the baryons containing two bottom quarks ($\Xi _{bb}$
and $\Omega _{bb}$ families) our calculations predict the lowest baryon mass
values. Predictions of Refs.~\cite{25GKLO00,26KLPS02} are also very close to
ours. The largest masses in this case (about 240--280~MeV higher then ours)
are given by \cite{19RP07}. For the triply-heavy baryons containing one
bottom and two charmed quarks ($\Omega _{bcc}$ family) Ref.~\cite{19RP07}
predicts the baryon masses 260~MeV higher than ours, the result similar as
in the case of $\Xi _{bb}$ and $\Omega _{bb}$ baryons. Predictions for the
masses of the triply-heavy baryons containing two bottom and one charmed
quark ($\Omega _{bbc}$ family) in our model and in Refs.~\cite
{25GKLO00,26KLPS02} are similar again, and mass values from Ref.~%
\cite{19RP07} are commonly higher than ours (in this case by about 400~MeV). The
largest difference between predictions of Ref.~\cite{19RP07} and baryon mass
values calculated in our model (560 MeV) is obtained for the heaviest ground
state baryon $\Omega _{bbb}$. As regards the predictions for the baryon mass
splittings, the situation in the bottom sector is evidently simpler, and
almost all approaches give similar results. The reason is also clear: the
interaction in this case is much weaker and consequently it causes smaller mass
splittings. Because the heavier systems seem to be simpler, one could
naively expect that in such case all reasonable approaches give similar
predictions. As we have seen from Tables~\ref{t3.6} and \ref{t3.7}, in
general this is not the case. Moreover, the situation with theoretical
predictions of the heavy baryon masses seems to be controversial to some
extent. For the baryons containing only one heavy quark all approaches, as a
rule, give similar predictions in reasonable agreement with available
experimental data. On the other hand, in the case of doubly-heavy (as well
as triply-heavy) baryons the predicted values strongly depend on the model.
Sometimes very different approaches (e.g. Refs.~\cite{25GKLO00,26KLPS02}
and our modified bag model) give very similar predictions, however, this is
rather an exception than a rule. Evidently, further investigations in this
field are necessary. On the other hand, the rapid development of
experimental and theoretical methods in recent years is rather impressive,
and we expect that in the nearest future new experimental data and new
improved results of lattice calculations will shed some light on the subject.

\section{Conclusions and discussion}

In conclusion, we have calculated the spectrum of all ground state baryons
containing bottom quarks by means of a modified bag model suitable for the
unified description of heavy and light hadrons. The model parameters are
practically the same as in our previous paper~\cite{39BS04} (following the
procedure proposed in the original MIT bag version the main four of them
have been determined from the light hadrons). For completeness the
calculated masses for the hadrons belonging to the charm sector are also
presented. All the predictions are compared with the calculations in other
approaches and with experimental data where available. For mesons and
baryons containing one heavy quark the agreement is good. Therefore we
expect our predictions for the doubly-heavy (and, maybe, triply-heavy)
baryon masses to be useful complementary tool in the treatment of heavy
baryon spectra.

Regretfully, the predictional power of the bag model is not so high, as, for
example, potential model in its various incarnations. The annoying thing is
that the applications of the bag model practically are restricted to the
calculation of properties of just the ground state hadrons. There was some
work done to incorporate the excited states in various versions (usual MIT
bag, chiral bag, etc.) of the model \cite{70D76,71MW84,72UM89,73OH99}. In
general the attempts were rather successful, however, the complications
associated with relativity and the problem of spurious center-of-mass motion
makes the bag model in this case calculationally much more unwieldy than the
nonrelativistic models. On the other hand, in some cases the initial
simplicity of the model may be regarded as an advantage, and old-good bag
model could serve for a while as a modest but still useful tool for
investigation of various hadronic properties (ground state hadron masses,
magnetic moments, isospin splittings, etc.), especially when the preliminary
quick estimate is necessary.

\end{document}